\documentclass[conference]{IEEEtran}
\IEEEoverridecommandlockouts
\usepackage{cite}
\usepackage{amsmath,amssymb,amsfonts}
\usepackage{algorithmic}
\usepackage{graphicx}
\usepackage{textcomp}
\usepackage{xcolor}
\usepackage{amsmath}
\usepackage{subcaption} % This package provides the \subcaption command

\def\BibTeX{{\rm B\kern-.05em{\sc i\kern-.025em b}\kern-.08em
    T\kern-.1667em\lower.7ex\hbox{E}\kern-.125emX}}
\begin{document}

\title{Optimizing Underwater IoT Routing with Multi-Criteria Decision Making and Uncertainty Weights\\

}

\author{
\IEEEauthorblockN{1\textsuperscript{rd} Ali Karkehabadi}
\IEEEauthorblockA{\textit{Electrical and Computer Engineering} \\
\textit{University of California, Davis}\\
Davis, USA \\
akarkehabadi@ucdavis.edu}

\and
\IEEEauthorblockN{2\textsuperscript{nd} Mitra Bakhshi}
\IEEEauthorblockA{\textit{Department of Civil Engineering} \\
\textit{Sapienza University of Rome}\\
Rome, Italy \\
bakhshi.2058909@studenti.uniroma1.it}

\and
\IEEEauthorblockN{3\textsuperscript{st} Seyed Behnam Razavian}
\textit{Arya institute}\\
Istanbul, Turkey \\
behnamrazavian@ieee.org}

\maketitle

\begin{abstract}

Effective data routing is vital in the Internet of Things (IoT) paradigm, especially in underwater mobile sensor networks where inefficiency can lead to significant resource consumption. This article presents an innovative method designed to enhance network performance and reduce resource usage, while also accurately determining component weights in these networks, ensuring quality service. Building upon previous research on multi-criteria decision-making systems in coastal RPL networks, our method involves key adaptations for underwater environments. It integrates comprehensive network features to identify the optimal parent node for each sensor, employing the fuzzy SWARA decision-making approach under uncertain conditions. This method takes into account various factors including hops, energy, ARSSI rate, delay, ETX, link delivery rate, and depth to determine the most effective parent node assignment. Through simulation, our approach demonstrates marked improvements in network performance compared to existing solutions. These advancements are significant, offering a new direction in enhancing underwater IoT communications and suggesting wider applications for IoT systems facing similar challenges.

\end{abstract}


\section{Introduction}

The Internet of Underwater Things (IoUT) has revolutionized the way underwater devices and sensors communicate and exchange data. A key protocol in this advancement is the Routing Protocol for Low-Power and Lossy Networks (RPL), which is specifically designed for low-power and resource-constrained devices in the IoUT environment. Its significance has been emphasized in recent studies \cite{monika2022efficient}. The proliferation of IoT devices has made underwater acoustic sensor networks increasingly important for marine scientists and enterprises, as noted in \cite{bello2022internet}. These networks, as part of the broader underwater IoT challenge, include stationary and mobile sensors/actuators and autonomous underwater vehicles, highlighting the diverse applications in military, environmental, and industrial fields \cite{jiang2023survey,zhu2022adaptive,gola2020underwater}. Historically, the focus of underwater communication research was on physical layer challenges and signal processing. However, the expansion of underwater networks, driven by the need for applications like underwater environment monitoring, has emphasized the importance of networking and analyzing sensor outputs \cite{jahanbakht2021internet,jung2021iot}. These networks, due to the unique challenges of the underwater environment, require resilience and stability that surpass terrestrial networks, especially given the harsh conditions underwater.

Underwater Wireless Sensor Networks (UWSNs) have become instrumental in the IoUT, comprising nodes with sensors communicating through acoustic waves. The nodes in these networks, which can be stationary or mobile, relay data to coastal stations via buoyant gateway nodes. This makes routing protocols crucial for optimal network performance \cite{gola2023empirical}. There are two primary categories of protocols for underwater sensor networks: location-dependent and location-independent. Location-based routing protocols often face challenges due to the unpredictable nature of marine environments and the limited effectiveness of GPS technology underwater \cite{acevedo2021wrf}. Energy efficiency is a key consideration in these protocols, as the nodes generally rely on non-rechargeable batteries \cite{bello2022internet, hussain2023cr, jiang2023survey}. The development of routing protocols for the IoUT involves addressing several unique challenges of the aquatic environment. Energy conservation is paramount since underwater networks cannot rely on solar charging, and battery replacement is complex. Effective routing systems must distribute network resources evenly to avoid overloads and ensure longevity \cite{lin2022decentralized,jung2021iot}. The growth of UWSN is driven by the increased interest in underwater environment monitoring for various purposes. The global Wireless Sensor Networks market is expected to grow significantly, with underwater networks being a key contributor \cite{yisa2021security}. A major challenge in underwater routing is the unreliability of GPS data for nodes and neighboring devices. GPS is often ineffective below certain depths, and transmitting node information based on depth and neighbor lists remains a challenge \cite{gola2023empirical,jung2021iot}. Traditional RPL implementations may not be sufficient for dynamic underwater conditions, leading to the proposal of integrating uncertain weights into RPL using SWARA and fuzzy SWARA MCDM techniques, thus enhancing its effectiveness in IoUT \cite{sathish2023review, prajapati2019prioritizing}.

This work adapts the terrestrial RPL protocol, originally developed by the IETF Group for IoT and LLN networks, for marine applications. It includes modifications across physical, data, network, and transmission layers, and incorporates node mobility in simulations.

\section{Theoretical Background}

\begin{figure}[h]%
\centering
\begin{minipage}{0.23\textwidth}
  \centering
  \includegraphics[width=0.8\linewidth]{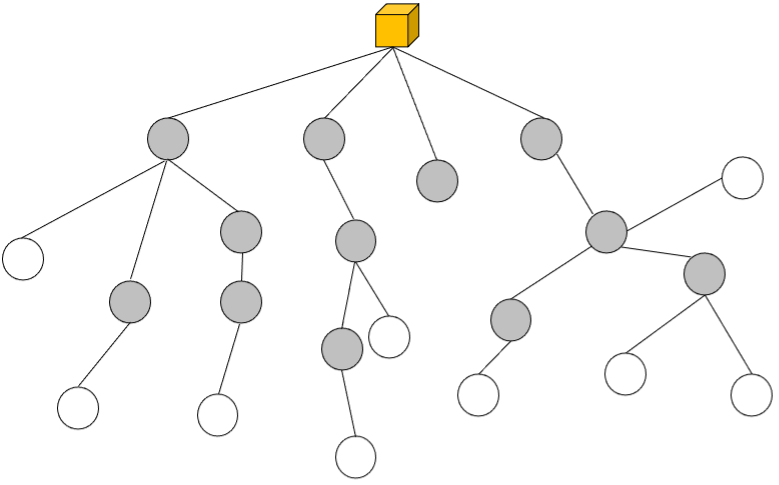}
  %\captionsetup{font=small} % Set subcaption font size to small
  \subcaption{Formed graph with limit degree $\gamma = 1$}
\end{minipage}%
\hfill
\begin{minipage}{0.23\textwidth}
  \centering
  \includegraphics[width=0.8\linewidth]{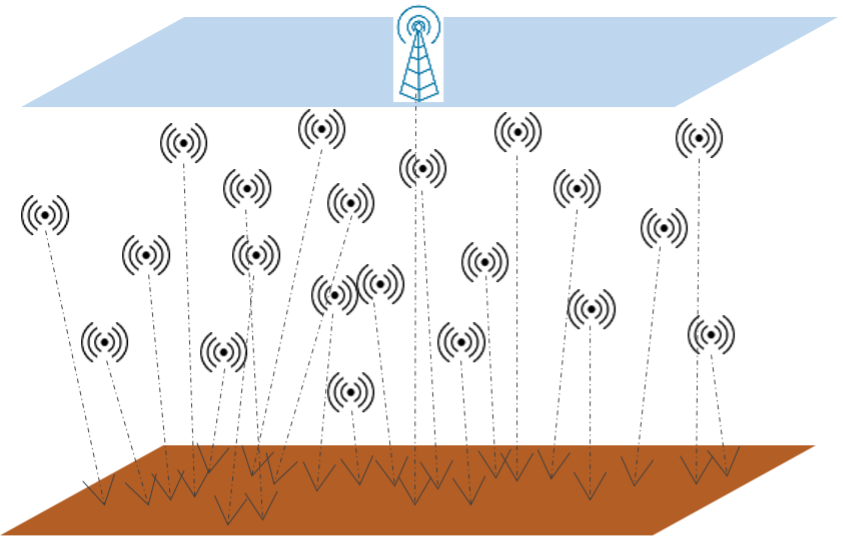}
  %\captionsetup{font=small} % Set subcaption font size to small
  \subcaption{A network with several underwater sensors}
  \label{Example_prl}
\end{minipage}
\caption{An example of an RPL network graph formation method in an underwater environment without parental degree limitation}
\label{example of an RPL network}
\end{figure}

The Routing Protocol for Low-Power and Lossy Networks (RPL) is designed for Wireless Sensor Networks (WSN) and Low-Power and Lossy Networks (LLN) within the Internet of Things (IoT) framework. Adapting RPL for underwater sensor networks requires modifications to address the differences between terrestrial wireless sensor networks and underwater acoustic networks. Optimizing underwater IoT routing involves designing efficient communication paths for submerged devices, and addressing challenges like limited bandwidth and high latency (1). Researchers are investigating techniques such as acoustic communication and energy-efficient routing, pivotal for successful deployments (2,3).
An example of an RPL network graph formation method in an underwater environment without parental degree limitation as illustrated in Figure \ref{example of an RPL network}. these adaptations, detailed in subsections 3.1 to 3.5, involve structural changes at various network levels and have been implemented using RPL code in the NS2 environment, with the integration of the Aquasim package.

\subsection{Speed of Sound in Water}

The underwater speed of sound (\( \Upsilon \)) varies with water temperature (\( T \)), salinity (\( \psi \)), and depth (\( d \)). Mackenzie's formula provides an accurate estimation  \cite{hussain2023cr, khan2022comprehensive, sathish2023review}:
\begin{align}
\Upsilon = & 1349 + 4.601T - 5.303 \times 10^{-2} T^2 + 2.369 \times 10^{-4} T^3 \nonumber \\
& + 1.35 (\psi - 35) + 1.64 \times 10^{-2} d + 1.685 \times 10^{-7} d^2 \nonumber \\
& + 1.026 \times 10^{-2} T(\psi-35) - 7.140 \times 10^{-3} Td^3
\end{align}

\subsection{Underwater Frequency Link Quality Criterion}

Underwater communication link quality is impacted by various noise sources like turbulence \(N_t\), boat movement \(N_s\), and wind-generated waves \(N_w\), along with ambient thermal noise \(N_{th}\). The spectral density power \(N(f)\) integrates these noises:
\begin{equation}
N(f) = N_t(f) + N_s(f) + N_w(f) + N_{th}(f)
\end{equation}

The signal-to-noise ratio (SNR) and communication channel capacity \(C(d,f)\) are calculated as follows, considering the increased noise rate reduces channel capacity:
\begin{equation}
\text{SNR} = \frac{P}{A(d,f)N(f)}
\end{equation}
\begin{equation}
C(d,f) = B \log_2(1 + \text{SNR}(d,f))
\end{equation}

\subsection{Delay Time Model}
Data transmission in underwater sensor networks uses sound waves, resulting in significant propagation delays. The total delay \(DT_{\text{all}}\) includes processing, queuing, propagation, and transmission delays.\cite{bello2022internet, jung2021iot, monika2022efficient}:

\begin{equation}
DT_{\text{all}} = \sum_{i=1}^{N} (T_{\text{proc}} + T_{\text{queue}} + T_{\text{prop}} + T_{\text{trans}})
\end{equation}

\subsection{Calculating Node Depth}

Each node's depth is measured using a pressure-based depth gauge. The depth difference \(\Delta d\) between two nodes is calculated as:
\begin{equation}
\Delta d = d_A - d_B = \frac{P_1}{\rho g} - \frac{P_2}{\rho g}
\end{equation}

\subsection{Frequency Attenuation or Absorption Model}
The absorption rate \(\alpha\), in dB/km, is influenced by factors like water depth, temperature, salinity, and pH. The frequency-dependent attenuation is modeled as follows, with specific assumptions for \(T\), \(d\), \(\Psi\), and \(pH\):
\begin{align}
\alpha = & 0.111 \frac{f_1 f^2}{f_1^2 + f^2} e^{\frac{(pH-8)}{0.56}} + 
0.52 \frac{1+T}{43} \frac{\Psi}{25} \frac{f_2 f^2}{f_2^2 + f^2} e^{-\frac{d}{6}} \nonumber\\
& + 5 \times 10^{-4} f^2 e^{-\frac{T}{27} - \frac{d}{17}}
\end{align}

\section{The Proposed RPLUW Method}
\subsection{System Model}

\begin{figure}[h]%
\centering
\begin{minipage}{0.49\textwidth}
  \centering
  \includegraphics[width=\linewidth]{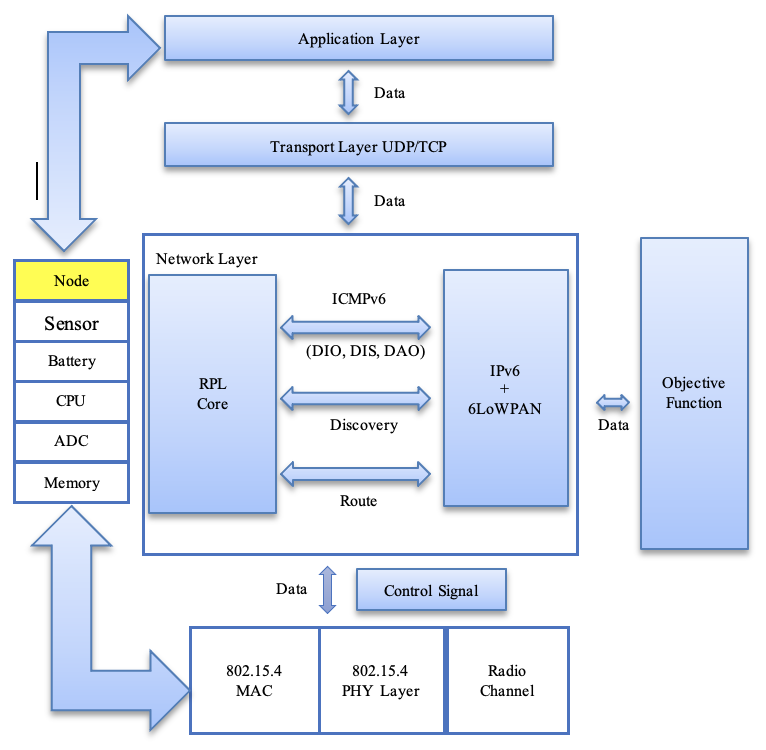}
  %\subcaption{Fixed Nodes}\label{subfig1}
\end{minipage}%
\hfill
\begin{minipage}{0.49\textwidth}
  \centering
  \includegraphics[width=\linewidth]{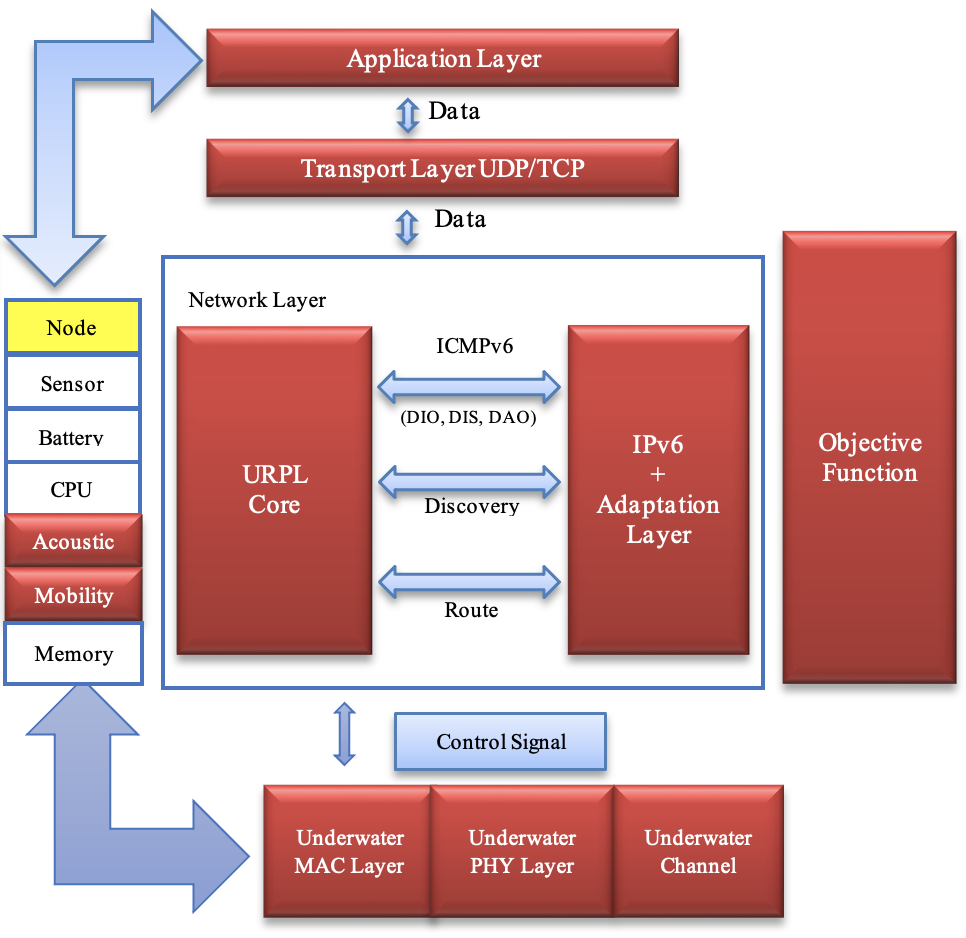}
  %\subcaption{Mobile Nodes}\label{subfig2}
\end{minipage}
\caption{RPL protocol structure in two normal sensor network environments and underwater sensor}
\label{RPL}
\end{figure}

The RPL is crucial in sensor network architectures, particularly in environments with limited power and bandwidth. This study, as illustrated in Figure \ref{RPL}, introduces two enhancements to RPL: URPL for stationary networks and RPLUW for underwater sensor network mobility. Key Components of the Proposed Protocol \cite{gola2023empirical, jiang2023survey, khan2022comprehensive, lilhore2022depth, sathish2022performance}:
\begin{itemize}
    \item \textbf{Rank \( R(u, j) \)}: A metric to evaluate the logical distance of a node \( u \) to the network graph root \( J \), adapting the RPL's Objective Function (OF) for underwater network specifics like node depth.
    
    \item \textbf{Preferred Parent DODAG (DPP)}: For a node \( u \) in graph \( G \), \( N(u) \) denotes its one-hop neighbors. The subset \( DPP(u,j) \) includes neighbors with the lowest rank \( R(u, j) \) to root in DODAG \( j \), from the set \( B \), allowing for multiple preferred parents and optimizing packet transmission paths.
    
    \item \textbf{DODAG Root List in DRL}: Utilizing Deep Reinforcement Learning (DRL), each node \( v \) in \( N(u) \) sends DIO packets with DODAG root locations. Thus, \( DRL(j) \) is an array at each node \( u \) storing these root locations, enhancing network adaptability.
\end{itemize}

\section{Handling Node Displacement in Underwater RPL Networks}

Node displacement in underwater RPL networks leads to graph instability, as mobile nodes frequently alter their neighbor lists. To manage this, two effective methods are employed:

\subsection{Efficient Trickle Timer Mechanism}

The Trickle Timer algorithm is designed to balance control packet transmissions and network graph stability. Operating within time intervals \([I_{\text{min}}, I_{\text{max}}]\), the method adapts the interval based on network stability. The interval doubles when the network is stable and resets to \(I_{\text{min}}\) when incompatibility count (IC) exceeds the threshold ``k'', promoting dynamic network topology updates. This approach effectively mitigates the impacts of node relocation, ensuring efficient network maintenance.

\subsection{Optimized IPv6 Neighbor Discovery}

To adapt to environmental shifts, RPL implements an optimized Neighbor Discovery (ND) process, utilizing ICMPv6 control packets for effective neighbor management. This includes:

\begin{itemize}
    \item \textbf{Neighbor Solicitation and Advertisement:} For confirming neighbor availability and communicating link changes.
    \item \textbf{Router Solicitation and Advertisement:} Where hosts request and routers broadcast essential network information, facilitating efficient graph adjustments.
\end{itemize}

\subsection{Timers in the RPLUW Method}

RPLUW integrates several timers to enhance underwater network reliability and stability \cite{khan2022comprehensive, lin2022decentralized, zhu2022adaptive, shwetha2024systematic}:

\begin{enumerate}
    \item \textbf{Linkage Timer (Lt):} Monitors channels and initiates discovery when parent nodes disconnect.
    \item \textbf{Mobility Timer (Mt):} Evaluates link reliability post DIO packet reception and facilitates node re-parenting.
    \item \textbf{Response Timer (Rt):} Ensures prompt parent node responses to DIS packets to minimize packet collisions and network disruptions.
\end{enumerate}

\begin{figure}[hpbt!]
\centering
\includegraphics[width=0.49\textwidth]{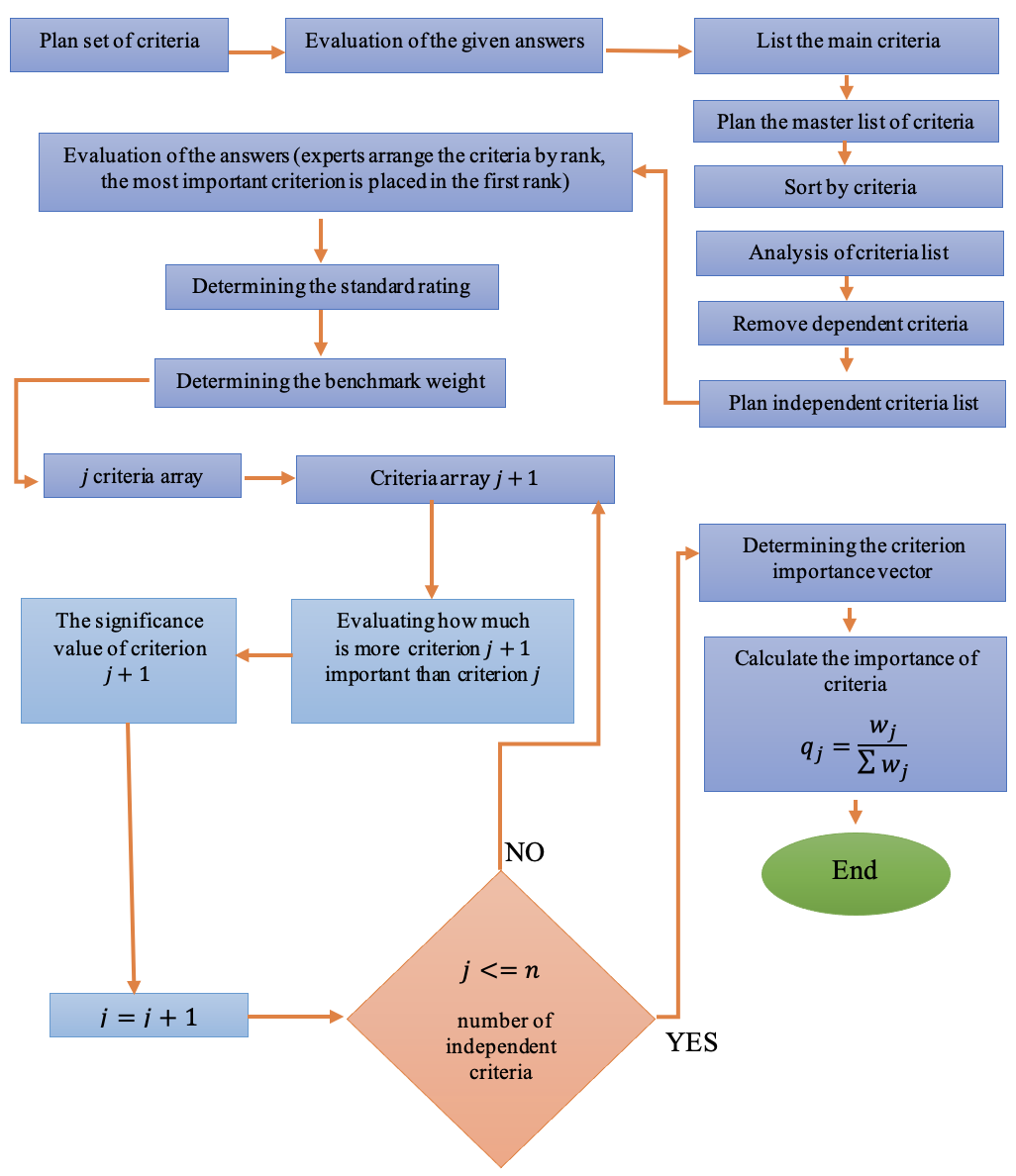}
\caption{Framework of Methodology}\label{fig2}
\end{figure}

\section{Research Methodology}
\subsection{RPLUW Network Graph Construction Method}

RPLUW, tailored for underwater IoT, uses Multi-Criteria Decision Making (MCDM) with uncertainty weighting. It considers factors like link quality, energy usage, battery life, and node trust. Nodes share link and energy data, enabling adaptive responses to underwater conditions. Figure \ref{fig2} illustrates our methodology.
The graph construction steps are:

\begin{enumerate}
    \item Sink broadcasts DIO packet.
    \item Nodes reply with DAO packets, sharing the depth and ARSSI rate.
    \item Sink confirms with DAO-Ack to first-level nodes.
    \item Parent nodes relay DIO packets, updating hop values, depth, and ARSSI.
    \item Nodes form potential parent lists from DIO packets.
    \item Child nodes request DAO membership from potential parents.
    \item Parents approve with DAO-Ack if the link quality is sufficient.
    \item Continues until the network graph is complete.
\end{enumerate}

\subsection{Stepwise Weight Assessment Ratio Analysis (SWARA)}

The SWARA method developed by Kersuliene, Zavadskas, and Turskis in 2010, SWARA assesses attribute weights based on expert judgment. Used in various applications, it calculates the relative importance \( S_j \) of attributes from decision-maker evaluations. SWARA enables experts to assess criteria and determine their weights collaboratively, enhancing decision-making by allowing for flexibility in estimating importance ratios. Its simplicity makes it accessible even to non-experts, providing a structured approach for well-informed decisions across domains \cite{mahdiyar2020barriers}. We utilize Fuzzy SWARA to address this, leveraging Fuzzy Logic's flexibility in handling ambiguous situations and imprecise data. Fuzzy Logic's benefits extend to accommodating partial truths, many-valued logic, and various applications, making it valuable for navigating uncertainty and informed decision-making \cite{sabetahd2022response}. Initially, decision-makers rank attributes by priority \cite{mahdiyar2020barriers, prajapati2019prioritizing}.\\
The coefficient \( K \) for an attribute is defined as:
\begin{equation}
\small
K_j = 
\begin{cases} 
1 & \text{if } j=1 \\
S_{j+1} & \text{if } j>1 
\end{cases}; \quad j=1,\dots,n
\end{equation}

The primary weight of an attribute is computed using:
\begin{equation}
\small
q_j = 
\begin{cases} 
1 & \text{if } j=1 \\
q_j/K_j & \text{if } j>1 
\end{cases}; \quad j=1,\dots,n
\end{equation}

Finally, to assign a weight to an attribute, the Relative Weight is calculated as follows:
\begin{equation}
\small
w_j = \frac{q_j}{\sum_{j=1}^n q_j}
\end{equation}

\begin{table}
\centering
\small
\caption{Statistical Analysis for the Problem Measures.}
\begin{tabular}{ccccccc}

\textbf{Criteria} & \textbf{Mean} & \textbf{STD} & \textbf{Min} & \textbf{Max} & \textbf{Median} \\

C1: Energy & 2.68 & 1.31 & 0 & 5 & 4\\
C2: Depth & 3.05 & 1.91 & 0 & 4 & 4\\
C3: Hop Count & 2.55 & 1.64 & 0 & 5 & 3\\
C4: Link Metrics & 2.82 & 1.40 & 0 & 4 & 3\\
C5: Other Metrics & 3.12 & 1.23 & 0 & 5 & 2\

\label{t1}
\end{tabular}
\end{table}

\begin{table}
\centering
\small
\caption{The final weight of the criteria}
\begin{tabular}{cccc}

\textbf{Criteria} & \textbf{Attribute Ranking} & \textbf{Kj} & \textbf{Initial weight} \\

C1: Energy & 0.246 & 1.138 & 0.086 \\
C2: Depth & 0.453 & 1.316 & 0.159 \\
C3: Hop Count & 0.840 & 1.916 & 0.291 \\
C4: Link Metrics & 0.98 & 1.151 & 0.343 \\
C5: Other Metrics & 0.345 & 1.291 & 0.121 \\
Sum & & &   1\\

\label{t2}
\end{tabular}
\end{table}

\begin{figure}[hpbt!]
\centering
\includegraphics[width=0.4\textwidth]{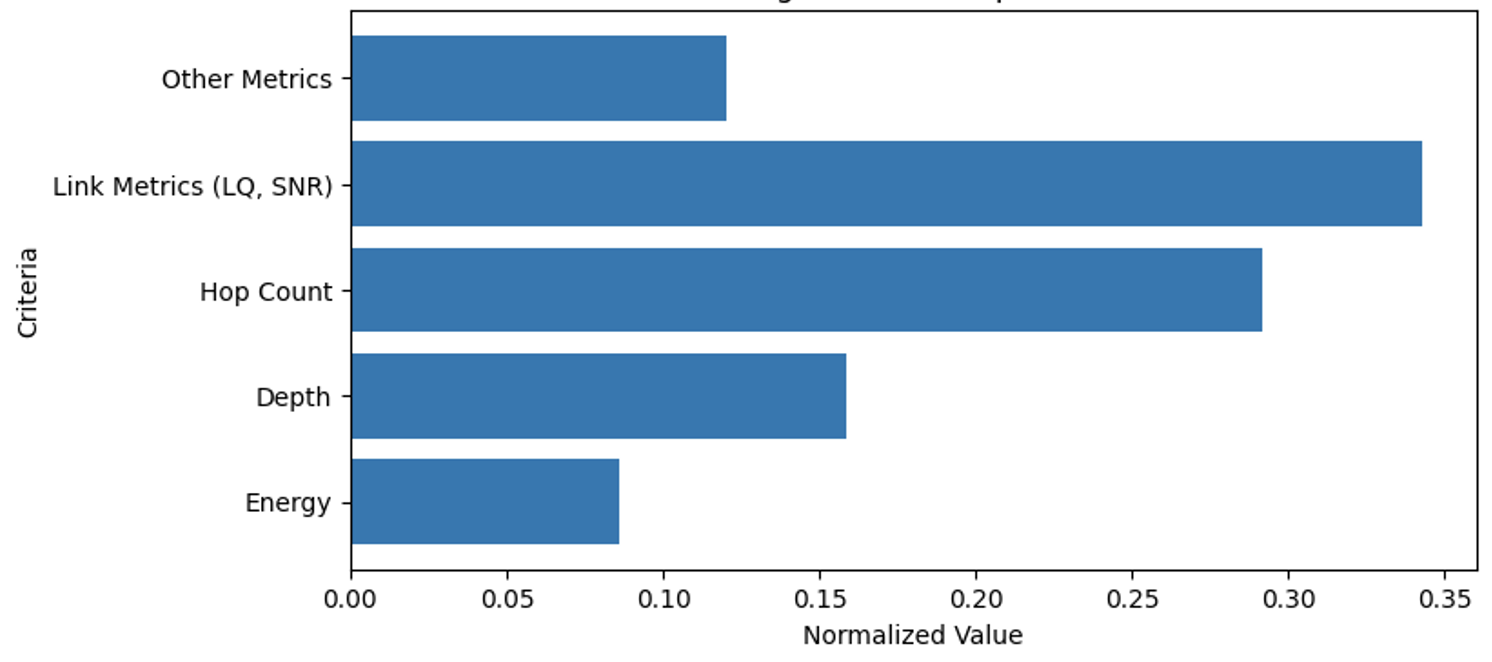}
\caption{Average end-to-end latency of network packets in the case of fixed and mobile nodes}\label{fig3}
\end{figure}

\begin{figure}[hpbt!]
\centering
\includegraphics[width=0.4\textwidth]{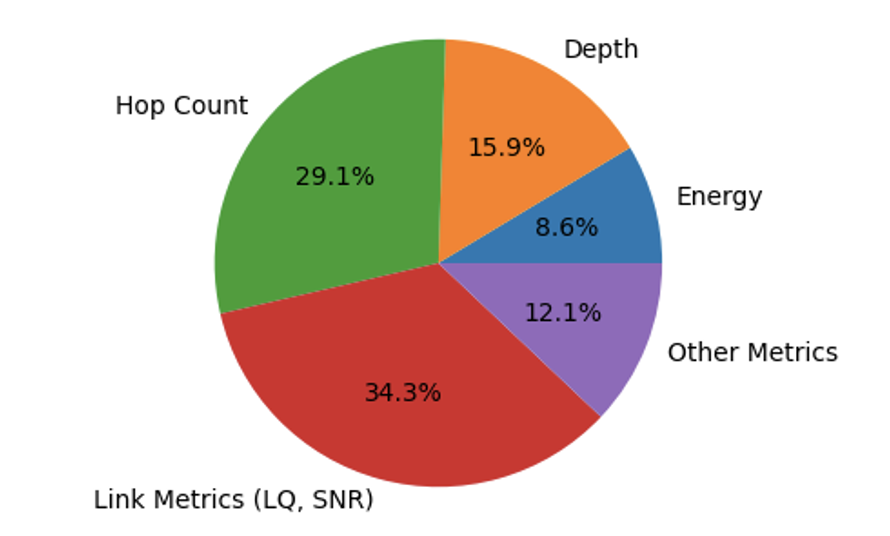}
\caption{Average end-to-end latency of network packets in the case of fixed and mobile nodes}\label{fig4}
\end{figure}

The tables \ref{t1} present statistical analysis, whereas \ref{t2} displays the final weights of the criteria. Figures \ref{fig3} and \ref{fig4} depict the normalized weight results for the Next Hop Selection Criteria.


\subsection{Routing in the RPLUW Network}

Routing is vital for data transmission in IoT. Multi-path techniques enhance network robustness by avoiding bottlenecks. This section proposes a method to boost efficiency and reduce resource use in underwater sensor networks while maintaining quality of service. Previous studies introduced a multi-criteria decision system for RPL networks offshore. Adapting it for underwater networks requires modifications in routing. This system integrates network features to find optimal parent nodes for sensor nodes. Each node maintains a list of potential parents limited by $\kappa = 4$ to avoid congestion and complexity.

The SWARA multi-criteria decision-making methodology assists in prioritizing parent nodes, aiming for optimal selection by child nodes. The weighting of parent nodes incorporates several attributes, including hop, energy, ARSSI rate, latency, ETX, link delivery rate, and depth. These factors collectively determine the final value assigned to each parent node.
\begin{align*}
\left[\begin{array}{c}
\omega_1 \\
\omega_2 \\
\omega_3 \\
\omega_4 \\
\omega_5 \\
\end{array}\right]
& \rightarrow \left[\begin{array}{c}
0.086279 \\
0.158697 \\
0.291407 \\
0.343035 \\
0.120582 \\
\end{array}\right] 
& \text{(a) Weights via SWARA}
\end{align*}
\begin{align*}
\left[\begin{array}{c}
\omega_1 \\
\omega_2 \\
\omega_3 \\
\omega_4 \\
\omega_5 \\
\end{array}\right]
& \rightarrow \left[\begin{array}{c}
0.088090 \\
0.162029 \\
0.296526 \\
0.330238 \\
0.123114 \\
\end{array}\right] 
& \text{(b) Weights via fuzzy SWARA}
\end{align*}

% \textbf{Figure 6:} Calculation of a multi-criteria decision system.

Post calculation, each child node's parent values list is refreshed. For mobile nodes, the ARSSI value is particularly updated at a heightened sampling rate. Equations (11) and (12) elucidate the decision system's calculations:

\begin{equation}
\small
\text{NodeValue}(k=1\dots n) = \sum_{k=1}^{n} \left( \text{Param}_k \times \omega_k \right)
\end{equation}
\begin{equation}
\small
\textbf{MADM}\text{ selection} = \textbf{MAX}(\text{NodeValue}_k)
\end{equation}
Subsequent sections will simulate and benchmark the proposed methodology against contemporary techniques. Table \ref{t3} illustrates the network simulation conditions.

\begin{table}[htbp]
\centering
% \footnotesize % Use an even smaller font size
\small
\caption{Network simulation conditions}
\begin{tabular}{|l|l|}
\hline
\textbf{Parameters} & \textbf{Value (s)} \\
\hline
Network topology & Random position\\
Frequency & 30.5 kHz \\
Deployment area & 1000$\times$1000$\times$500m$^3$\\
Channel & Underwater channel \\
Initial node energy & 50 J\\
Maximum Bandwidth & 30 kbps \\
Initial sink energy & 50 kJ\\
Packet size & 50 bytes \\
Number of nodes & 50, 100, 200\\
DIO packet size & 4 bytes \\
Nodes mobility & 1 m/s--5 m/s\\
DAO packet size & 4 bytes \\
Mobility model & Random mobility\\
DAO-Ack packet size & 4 bytes \\
Percentage of Mobile Nodes & 40\%\\
DIS packet size & 4 bytes \\
Cost of long transmission & 1.3 W \\
Packet generation rate & $\lambda = 0.1\text{--}0.2$ pkt/s \\
Cost of short transmission & 0.8 W\\
Memory size & 12 MB \\
Cost of reception & 0.7 W\\
Sink position & Surface(500 $\times$ 500 $\times$ 0) \\
Idle power & 0.008 W\\
Antenna & Omni-directional \\
Data aggregation power & 0.22 W\\
Simulation time & 600 \\
Communication range of ASN & 150 m\\
Iterations & 10 \\
Acoustic transmission range & 200 m\\
Number of Channels & (30.511 to 30.581) kHz \\
Spreading values & 1.3 \\
\hline
\end{tabular}
\label{t3}
\end{table}

\section{Enhanced System Model and Performance Evaluation}

\begin{figure}[h]%
\centering
\begin{minipage}{0.24\textwidth}
  \centering
  \includegraphics[width= \columnwidth, height=1.4 in]{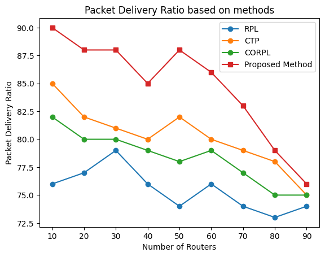}
  %\subcaption{Fixed Nodes}\label{subfig1}
\end{minipage}%
\hfill
\begin{minipage}{0.24\textwidth}
  \centering
  \includegraphics[width=\columnwidth, height= 1.4 in]{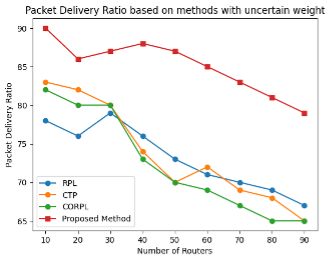}
  %\subcaption{Mobile Nodes}\label{subfig2}
\end{minipage}
\vspace{-5pt}
\caption{ comparison of packet delivery ratio (PDR) with number of routers}\label{fig5}
\end{figure}

Figure \ref{fig5} illustrates a comparative assessment of packet delivery ratios (PDR) among various algorithms in relation to router count. Our proposed algorithm achieves a notable 18.43\% surge in PDR, surpassing other existing algorithms. This improvement signifies reduced radio channel collisions and enhanced routing efficiency between routers and gateways. The algorithm's effectiveness lies in discerning the optimal gateway node for data intake, resulting in minimized control message travel distance and overall count, showcasing the efficacy of the proposed system.

\begin{figure}[h]%
\centering
\begin{minipage}{0.24\textwidth}
  \centering
  \includegraphics[width=\columnwidth, height=1.4 in]{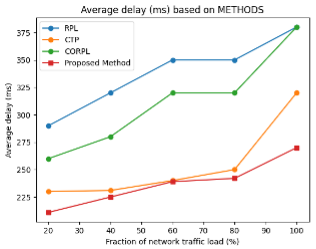}
  %\subcaption{Fixed Nodes}\label{subfig1}
\end{minipage}%
\hfill
\begin{minipage}{0.24\textwidth}
  \centering
  \includegraphics[width=\columnwidth, height=1.4 in]{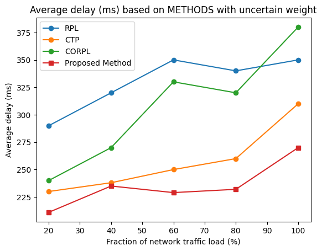}
  %\subcaption{Mobile Nodes}\label{subfig2}
\end{minipage}
\vspace{-5pt}
\caption{comparison of average delay and network traffic load rate}
\label{fig6}
\end{figure}

In Figure \ref{fig6}, we explore the correlation between average delay and network traffic load rate. Our algorithm demonstrates a significant 17.22\% reduction in average delay compared to its peers, enhancing load balancing among routers. Despite its iterative nature and sub-optimal solutions, it outperforms alternative strategies in terms of time complexity, improving PDR, throughput, and load balance across gateways.

\begin{figure}[h]
\centering
\includegraphics[width=0.49\textwidth]{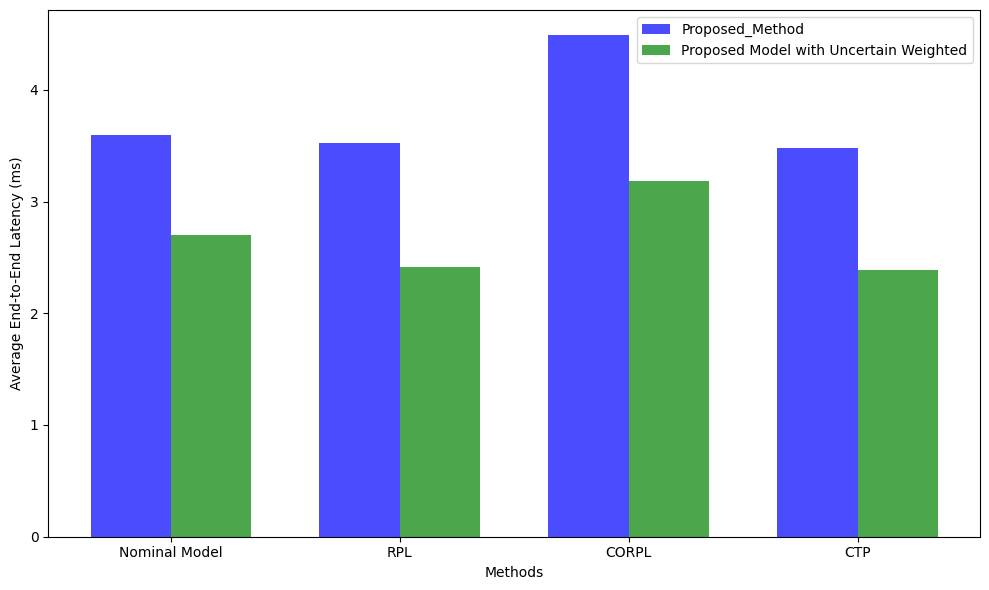}
\caption{Average end-to-end latency of network packets in the case of fixed and mobile nodes}
\label{fig7}
\end{figure}

Delving into another evaluation metric for underwater sensor networks, end-to-end latency emerges as a pivotal parameter. A network's success in environmental monitoring is often gauged by how reduced and consistent the average latency of its packets is. To put it differently, a downtrend in the network's Jitter rate can symbolize this metric. In our proposed model with uncertain weighted networks housing static nodes, the end-to-end latency, when juxtaposed with methods like CTP, CORPL, and RPL, witnesses an improvement oscillating between 3.48\% and 4.49\%. The mobile avatar of our approach, dubbed RPLUW with uncertain weight, showcases improvements ranging from 8\% to 11\%. This substantiates the efficacy of our multi-path routing, schedules, and judicious use of the decision system. Emphasizing the pivotal parameters during network link formulation and sustenance results in the longevity of the connections amongst nodes, outclassing many parallel methods. Figure \ref{fig7} captures the results of this analysis, both in static and dynamic node landscapes.\\

\section{Conclusion}

The progression in hardware technologies has catalyzed the feasibility of monitoring physical attributes via IoT networks, simultaneously fostering the advancement in communication network efficiency. In the realm of oceanic IoT applications, the need for robust platforms is evident, particularly those that enhance battery longevity and minimize response times for time-sensitive operations. Our study introduces a multi-criteria decision system, replete with a range of effective parameters, specifically designed for the orchestration, maintenance, and enhancement of network topology. Additionally, the intricate dynamics of mobile nodes in networks were elucidated using our proposed methodologies. By leveraging multi-route mechanisms, our approach adeptly modulates load equilibrium, thereby reducing traffic strain on network nodes. Decision systems facilitate topology regulation and network graph rejuvenation while ensuring computational parsimony. The outcomes of our simulations, when juxtaposed with contemporary methodologies, affirm the superior performance and efficiency of our proposed algorithm. Future research directions include the development of an energy integration model and a delay-mitigation framework based on our algorithm. Further, exploring learning systems to manage the complexities of data consolidation in subaqueous IoT networks presents a promising avenue for future research.


\begin{thebibliography}{00}

\bibitem{monika2022efficient}Monika, R., Dhanalakshmi, S., Kumar, R., Narayanamoorthi, R. \& Lai, K. An efficient adaptive compressive sensing technique for underwater image compression in IoUT. {\em Wireless Networks}. pp. 1-15 (2022)

\bibitem{bello2022internet}Bello, O. \& Zeadally, S. Internet of underwater things communication: Architecture, technologies, research challenges and future opportunities. {\em Ad Hoc Networks} pp. 102933 (2022)
\bibitem{gola2023empirical}Gola, K., Dhingra, M., Gupta, B. \& Rathore, R. An empirical study on underwater acoustic sensor networks based on localization and routing approaches. {\em Advances In Engineering Software}. pp. 103319 (2023)
\bibitem{gola2020underwater}Gola, K. \& Gupta, B. Underwater sensor networks:‘Comparative analysis on applications, deployment and routing techniques’. {\em IET Communications}. 2859-2870 (2020)
% \bibitem{halakarnimath2021location}Halakarnimath, B. \& Sutagundar, A. Location Estimation of Nodes in Underwater Acoustic Sensor Networks. {\em Journal Of Telecommunications And Information Technology}., 15-31 (2021)
\bibitem{hussain2023cr}Hussain, A., Hussain, T., Ullah, I., Muminov, B., Khan, M., Alfarraj, O. \& Gafar, A. CR-NBEER: Cooperative-Relay Neighboring-Based Energy Efficient Routing Protocol for Marine Underwater Sensor Networks. {\em Journal Of Marine Science And Engineering}. 1474 (2023)

\bibitem{jahanbakht2021internet}Jahanbakht, M., Xiang, W., Hanzo, L. \& Azghadi, M. Internet of underwater things and big marine data analytics—a comprehensive survey. {\em IEEE Communications Surveys}. (2021)

% \bibitem{jahanbakht2021internet}Jahanbakht, M., Xiang, W., Hanzo, L. \& Azghadi, M. Internet of underwater things and big marine data analytics—a comprehensive survey. {\em IEEE Communications Surveys & Tutorials}. 904-956 (2021)

\bibitem{jiang2023survey}Jiang, J., Han, G. \& Lin, C. A survey on opportunistic routing protocols in the Internet of Underwater Things. {\em Computer Networks}. \textbf{225} pp. 109658 (2023)
\bibitem{jung2021iot}Jung, L. IoT underwater wireless sensor network monitoring. {\em Role Of IoT In Green Energy Systems}. pp. 38-58 (2021)


\bibitem{khan2022comprehensive}Khan, Z., Gang, Q., Muhammad, A., Muzzammil, M., Khan, S., Affendi, M., Ali, G., Ullah, I. \& Khan, J. A comprehensive survey of energy-efficient MAC and routing protocols for underwater wireless sensor networks. {\em Electronics}. 3015 (2022)
% \bibitem{khandelval2021adaptive}Khandelval, M. \& Others An Adaptive Approach for Deploy Underwater Acoustic Network.  (2021)
\bibitem{lilhore2022depth}Lilhore, U., Khalaf, O., Simaiya, S., Tavera Romero, C., Abdulsahib, G. \& Kumar, D. A depth-controlled and energy-efficient routing protocol for underwater wireless sensor networks. {\em International Journal Of Distributed Sensor Networks}. 15501329221117118 (2022)
\bibitem{lin2022decentralized}Lin, B., Duan, J., Han, M. \& Cai, L. Decentralized Reinforcement Learning-Based Access Control for Energy Sustainable Underwater Acoustic Sub-Network of MWCN. {\em Next Generation Marine Wireless Communication Networks}. pp. 83-106 (2022)
\bibitem{mahdiyar2020barriers}Mahdiyar, A., Mohandes, S., Durdyev, S., Tabatabaee, S. \& Ismail, S. Barriers to green roof installation: An integrated fuzzy-based MCDM approach. {\em Journal Of Cleaner Production}. pp. 122365 (2020)
% \bibitem{mohammadi2023comprehensive}Mohammadi, R. A comprehensive Blockchain-oriented secure framework for SDN/Fog-based IoUT. {\em International Journal Of Information Security}. pp. 1-13 (2023)

\bibitem{prajapati2019prioritizing}Prajapati, H., Kant, R. \& Shankar, R. Prioritizing the solutions of reverse logistics implementation to mitigate its barriers: A hybrid modified SWARA and WASPAS approach. {\em Journal Of Cleaner Production}. pp. 118219 (2019)
\bibitem{sathish2022performance}Sathish, K., Ravikumar, C., Srinivasulu, A., Gupta, A. \& Others Performance analysis of underwater wireless sensor network by deploying FTP, CBR, and VBR as applications. {\em Journal Of Computer Networks And Communications}. (2022)

\bibitem{acevedo2021wrf}Acevedo, P., Jabba, D., Sanmartin, P., Valle, S., Nino-Ruiz, E. WRF-RPL: Weighted random forward RPL for high traffic and energy demanding scenarios.{\em IEEE Access}. (2021) 





\bibitem{yisa2021security}Yisa, A., Dargahi, T., Belguith, S. \& Hammoudeh, M. Security challenges of internet of underwater things: A systematic literature review. {\em Transactions On Emerging Telecommunications Technologies}. e4203 (2021)


\bibitem{zhu2022adaptive}Zhu, L., Yao, H., Chang, H., Tian, Q., Zhang, Q., Xin, X. \& Yu, F. Adaptive optics for orbital angular momentum-based internet of underwater things applications. {\em IEEE Internet Of Things Journal}. 24281-24299 (2022)
\bibitem{sathish2023review}Sathish, K., Venkata, R., Anbazhagan, R. \& Pau, G. Review of Localization and Clustering in USV and AUV for Underwater Wireless Sensor Networks. {\em Telecom}. 43-64 (2023)

% \bibitem{simon2024improved}Simon, J., Kapileswar, N., Phani Kumar, P. \& Aarthi Elaveini, M. Improved geographic opportunistic routing protocol for void hole elimination in underwater IoTs: Parameter tuning by TSA optimization. {\em International Journal Of Communication Systems}. \textbf{37}, e5659 (2024)
\bibitem{shwetha2024systematic}Shwetha, M. \& Krishnaveni, S. A Systematic Analysis, Outstanding Challenges, and Future Prospects for Routing Protocols and Machine Learning Algorithms in Underwater Wireless Acoustic Sensor Networks. {\em Journal Of Interconnection Networks}. pp. 2330001 (2024)

\bibitem{sabetahd2022response}Sabetahd, R., Mousavi Ghasemi, S., Vafaei Poursorkhabi, R., Mohammadzadeh, A. \& Zandi, Y. Response Attenuation of a Structure Equipped with ATMD under Seismic Excitations Using Methods of Online Simple Adaptive Controller and Online Adaptive Type-2 Neural-Fuzzy Controller. {\em Computational Intelligence And Neuroscience}. \textbf{2022}

\end{thebibliography}
\end{document}